\documentclass[11pt, letterpaper]{article}
\usepackage{osajnl}

\newcommand{\commentout}[1]{}

\usepackage{amsmath}
\usepackage{amssymb}

\newcommand{\nwc}{\newcommand}

\newcommand{\lt}{\left}
\nwc{\partz}{\frac{\partial }{\partial z}}
\newcommand{\rt}{\right}
\nwc{\ytil}{\tilde{\by}}
\nwc{\al}{\alpha}
\nwc{\half}{\frac{1}{2}}

\newcommand{\lan}{\left\langle}
\newcommand{\ran}{\right\rangle}

\newcommand{\bx}{\mathbf x}
\newcommand{\bR}{\mathfrak{R}}
\newcommand{\mbR}{\mathbf{R}}
\nwc{\fR}{\mathfrak{R}}
\newcommand{\bp}{\mathbf p}

\newcommand{\bq}{\mathbf q}
\newcommand{\by}{\mathbf y}

\newcommand{\cpv}{\!\!\!\!\!\! - \  }

\nwc{\nwt}{\newtheorem}

\nwt{proposition}{Proposition}
\nwt{lemma}{Lemma}
\nwt{theorem}{Theorem}

\nwt{remark}{Remark}
\nwt{assumption}{Assumption}
\nwt{definition}{Definition} 

\nwc{\bal}{\begin{align}}
\nwc{\be}{\begin{equation}}
\nwc{\ben}{\begin{equation*}}
\nwc{\bea}{\begin{eqnarray}}
\nwc{\beq}{\begin{eqnarray}}
\nwc{\bean}{\begin{eqnarray*}}
\nwc{\beqn}{\begin{eqnarray*}}
\nwc{\beqast}{\begin{eqnarray*}}

\nwc{\eal}{\end{align}}
\nwc{\ee}{\end{equation}}
\nwc{\een}{\end{equation*}}
\nwc{\eea}{\end{eqnarray}}
\nwc{\eeq}{\end{eqnarray}}
\nwc{\eean}{\end{eqnarray*}}
\nwc{\eeqn}{\end{eqnarray*}}
\nwc{\eeqast}{\end{eqnarray*}}

\nwc{\invf}{\cF^{-1}_2}
\nwc{\ep}{\ell}
\nwc{\tep}{\tilde{\varepsilon}}
\nwc{\epsq}{{\varepsilon^2}}
\nwc{\epsqa}{{\varepsilon^{2\alpha}}}
\nwc{\eps}{\varepsilon}
\nwc{\ept}{\epsilon}
\nwc{\vrho}{\varrho}
\nwc{\orho}{\bar\varrho}
\nwc{\ou}{\bar u}
\nwc{\vpsi}{\varpsi}
\nwc{\lamb}{\ell}

\nwc{\nn}{\nonumber}
\nwc{\bm}{\boldmath}
\nwc{\mf}{\mathbf}
\nwc{\mb}{\mathbf}
\nwc{\ml}{\mathcal}

\nwc{\IA}{\mathbb{A}} 
\nwc{\IB}{\mathbb{B}}
\nwc{\IC}{\mathbb{C}} 
\nwc{\ID}{\mathbb{D}} 
\nwc{\IM}{\mathbb{M}} 
\nwc{\IP}{\mathbb{P}} 
\nwc{\II}{\mathbb{I}} 
\nwc{\IE}{\mathbb{E}} 
\nwc{\IF}{\mathbb{F}} 
\nwc{\IG}{\mathbb{G}} 
\nwc{\IN}{\mathbb{N}} 
\nwc{\IQ}{\mathbb{Q}} 
\nwc{\IR}{\mathbb{R}} 
\nwc{\IT}{\mathbb{T}} 
\nwc{\IZ}{\mathbb{Z}} 

\nwc{\epal}{\ep^{-2\alpha}}
\nwc{\cE}{{\ml E}}
\nwc{\cP}{{\ml P}}
\nwc{\cQ}{{\ml Q}}
\nwc{\cL}{{\ml L}}
\nwc{\cR}{{\ml R}}
\nwc{\cV}{{\ml V}}
\nwc{\cT}{{\ml T}}
\nwc{\crV}{{\ml L}_{(\delta,\nu)}}
\nwc{\cC}{{\ml C}}
\nwc{\cA}{{\ml A}}
\nwc{\cK}{{\ml S}}
\nwc{\cB}{{\ml B}}
\nwc{\cD}{{\ml D}}
\nwc{\cF}{{\ml F}}
\nwc{\cS}{{\ml S}}
\nwc{\cM}{{\ml M}}
\nwc{\cG}{{\ml G}}
\nwc{\cH}{{\ml H}}
\nwc{\bk}{{\mb k}}
\nwc{\cbz}{\overline{\cB}_z}
\nwc{\bE}{{\mb E}}
\nwc{\bH}{{\mb H}}
\nwc{\bO}{{\mb O}}
\nwc{\bU}{{\mb U}}
\nwc{\bW}{{\mb W}}
\nwc{\bK}{{\mb K}}
\nwc{\bI}{{\mb I}}
\nwc{\bP}{{\mb P}}
\nwc{\bV}{{\mb V}}
\nwc{\bb}{{\mb e}}
\nwc{\bba}{{{\mb e}^{\sigma,\alpha}}}
\nwc{\bbz}{{{\mb e}^{\sigma,\zeta}}}
\nwc{\bbe}{{{\mb e}^{\sigma,\eta}}}
\nwc{\bB}{{\mb B}}
\nwc{\bEaz}{{{\mb E}^{\sigma,\alpha\zeta}}}
\nwc{\bF}{{\mb F}}
\nwc{\bG}{{\mb G}}
\nwc{\bT}{{\mb T}}
\nwc{\Waz}{{\bar W}^\sigma_{\alpha\zeta}}
\nwc{\Caz}{{C}^\sigma_{\alpha\zeta}}
\nwc{\bd}{{\mb d}}
\nwc{\bda}{{{\mb d}^{\sigma,\alpha}}}
\nwc{\bdz}{{{\mb d}^{\sigma,\zeta}}}
\nwc{\bde}{{{\mb d}^{\sigma,\eta}}}
\nwc{\bD}{{\mb D}}
\nwc{\bDaz}{{{\mb D}^{\sigma,\alpha\zeta}}}
\nwc{\Om}{{\Omega}}
\nwc{\bSig}{{\mb \Sigma}}
\nwc{\fM}{\mathfrak{M}}
\nwc{\bJ}{{\mb J}}
\nwc{\fH}{\mathfrak{H}}
\nwc{\fQ}{\mathfrak{P}}
\nwc{\ga}{\gamma}
\nwc{\bX}{{\mb X}}
\nwc{\fX}{\mathfrak{X}}
\nwc{\bY}{{\mb Y}}
\nwc{\fY}{\mathfrak{Y}}
\nwc{\fC}{\mathfrak{C}}
\nwc{\bn}{\mb n}
\nwc{\bu}{\mb u}
\nwc{\bg}{{\mb g}}
\nwc{\ppj}{\partial_{p_{j}}}
\nwc{\pxj}{\partial_{x_{j}}}
\nwc{\pxjtil}{\partial_{\tilde x_{j}}}
\nwc{\ppl}{\partial_{p_{l}}}
\nwc{\pxl}{\partial_{x_{l}}}
\nwc{\pxltil}{\partial_{\tilde x_{l}}}

\nwc{\pft}{\cF^{-1}_\bp}

\newcommand{\fW}{\mathfrak{W}}
\newcommand{\om}{\omega}
\nwc{\fS}{\mathfrak{S}}
\nwc{\fF}{\mathfrak{F}}
\nwc{\bQ}{{\mb P}}

\begin{document}

\title{Two-Frequency  Radiative Transfer.  II: Maxwell Equations in
Random Dielectrics}

\author{Albert C. Fannjiang
 \thanks{
The research is supported in part by the Defense Advanced Research Projects Agency (DARPA) grant 
 N00014-02-1-0603
}
}
\address{Department of Mathematics,
University of California,
Davis, CA 95616-8633}

\email{fannjiang@math.ucdavis.edu}
\begin{abstract}
The paper addresses  the space-frequency
correlations of electromagnetic waves in general
random, bi-anisotropic media  whose constitutive 
tensors  are complex Hermitian matrices. The two-frequency
Wigner distribution  (2f-WD) for polarized waves is introduced
to describe the space-frequency correlations
and the closed form Wigner-Moyal equation is
derived from the Maxwell equations. 
Two-frequency radiative transfer (2f-RT) equations
 are then derived  from the Wigner-Moyal equation by  using the multiple scale expansion. For the simplest isotropic medium, the result coincides with
Chandrasekhar's transfer equation. In birefringent media,
the 2f-RT equations take the scalar form due to the absence of
depolarization. A number of birefringent media 
 such as the chiral, uniaxial and gyrotropic  media are examined.
 For the unpolarized wave in the isotropic medium
 the 2f-RT equations reduces to the 2f-RT equation previously derived
 in Part I. A similar Fokker-Planck-type equation is derived
 from the scalar 2f-RT equation for the birefringent media. 
\end{abstract}


\maketitle

\section{Introduction}
In Part I \cite{2f-rt-josa} of the series we studied
the space-frequency correlation for {\em scalar} waves
in random media as governed  by the Helmholtz equation
with a randomly fluctuating  refractive index.
To this end,  we introduced the two-frequency
Wigner distribution (2f-WD) which in the unscaled form is 
\beqn
\label{1.1}
&&W(\bx,\bp;\omega_1, \omega_2)\\
&=&\frac{1}{(2\pi)^3}\int
e^{-i\bp^\dagger\by}
U_1 (\frac{\bx}{\omega_1}+
\frac{\by}{2\omega_1}){U^\dagger_2(\frac{\bx}{\omega_2}
-\frac{\by}{2\omega_2})}d\by
\eeqn
where $U_1$ and $U_2$ are the wave fields at
frequencies $\omega_1$ and $\omega_2$ respectively. 
Throughout, $\dagger$ denotes the Hermitian conjugation
of vectors or matrices. 
The important characteristic of the definition (\ref{1.1})
is that the spatial argument of each wave field is scaled
in proportion to the respective wavelength. The variables
$\bx$ are  
the so called size parameter in scattering theory when the
phase velocity is unity \cite{Mis}. 

In the weak coupling (disorder) regime we
derived
the two-frequency radiative transfer (2f-RT) equation for
the two-frequency Wigner distribution. We considered
several approximations, notably the geometrical
optics and paraxial approximations. Based on
the dimensional analysis of these asymptotic equations
we obtained scaling behavior of the coherence bandwidth
and coherence length. We also obtained the space-frequency correlation
{\em below} the transport mean-free-path by analytically solving
one of the paraxial 2f-RT equations.
 
The main advantage of the 2f-RT theory over the traditional equal-time RT theory is that it describes  not just the energetic transport  but also the two space-time point mutual coherence in the following way. 
Let $u(t_j,\bx_j), j=1,2$ be the time-dependent wave field
at two space-time points $(t_j, \bx_j), j=1,2.$ Let
$\bx=(\omega_1\bx_1+\omega_2\bx_2)/2$ and $\by=\omega_1\bx_1-\omega_2\bx_2$. Then we have
\beq
\label{sfc}
&&\lan u(t_1,\bx_1) u^*(t_2,\bx_2)\ran\\
&=&\int e^{i(\omega_2t_2-\omega_1t_1)} \lan U_1(\bx_1)
U_2^\dagger(\bx_2)\ran d \omega_1 d\omega_2\nn\\
&=& \int e^{i\bp^\dagger\by}  e^{-\om'
t}e^{-i\tau\om} \lan W(\bx,\bp; \om+\om'/2, \om-\om'/2)\ran
 d\om d\om' d\bp\nn
\eeq
with $t=(t_{1}+t_{2})/2, \tau=t_{1}-t_{2}, \om=(\om_1+\om_2)/2, \om'=\om_1-\om_2$. Here and below $\lan \cdot\ran$ is
the ensemble averaging w.r.t. the medium fluctuations,
$*$ the complex conjugation and $\dagger$ the Hermitian
conjugation. 
 In comparison, the single-time correlation gives rise to the expression 
\beqn
&&\lan u(t,\bx_1) u^\dagger(t,\bx_2)\ran\\
&=& \int  e^{i\bp^\dagger\by}e^{-i\om' t}\lt[\int \lan W(\bx,\bp; \om+\om'/2, \om-\om'/2)\ran
d \om \rt]d\om'd\bp
\eeqn
which, through spectral decomposition,  determines only  the central-frequency-integrated  2f WD. For a statistically stationary
signal,  (\ref{sfc}) would be a function of $t_1-t_{2}$
only. In this case  different frequency components are uncorrelated
and consequently  only the equal-frequency WD is necessary to describe the two-spacetime correlation \cite{MW}. 
For statistically non-stationary signals 
the two-frequency cross-correlation is needed 
to characterize the two-spacetime correlation. 

The 2f-RT theory developed in Part I
has enabled precise estimate  of important physical quantities  such
as the coherence length and the coherence bandwidth \cite{2f-rt-josa}  which 
are medium characteristics relevant to communications and
imaging in disordered media \cite{pulsa-nl,mfirm-rice}.
In particular, the two-frequency formulation  is
an indispensable tool for the statistical stability analysis of the time-reversal
communication scheme with
broadband signals in multiple-scattering media (see Ref. \cite{pulsa-nl} where a 2f-RT equation and its solution
play a key role).  The 2f-RT theory developed here is expected
to extend these results to  
the case of polarized waves. 

The organization of this paper is as follows. 
In Section~\ref{sec2} and Appendix A we develop the two-frequency
formulation of the Maxwell equations for general heterogeneous
dielectric in terms of 2f-WD.
In Section~\ref{sec3}, we formulate the weak-coupling scaling limit for two-frequency Wigner-Moyal equation.
In Section~\ref{sec4} we develop the multiscale expansion
to find an approximate solution in the weak-coupling regime.
In Section \ref{sec5} and Appendix B, based on 
a solvability condition we give an explicit
form to the 2f-RT equations 
for general bi-anisotropic media and in Section \ref{sec:scalar}
we derive a scalar 2f-RT equation for birefringent media.  In Section~\ref{sec6},
we consider the isotropic medium
and show that the general 2f-RT equations, after a change
of variable, reduces to the two-frequency version of
Chandrasekhar's transfer equation. 
In Section \ref{sec:chi},  \ref{sec:aniso} and \ref{sec:gyro}, we examine three birefringent media: the chiral, the uniaxial and the gyrotropic media. 
In Section~\ref{sec7} we analyze the unpolarized wave
in the isotropic medium 
in the geometrical optics regime and show that Chandrasekhar's
equation  reduces to a Fokker-Planck-type
equation rigorously derivable from the geometrical optics
of the scalar wave  
\cite{2f-grt}.
We derive a similar equation from the scalar
2f-RT equation for the birefringent media. 
We conclude the paper in Section~\ref{sec:conc} with
a brief discussion on expressing the two-spacetime correlation
in terms of solutions of the 2f-RT equations.

\section{Maxwell equations and Wigner-Moyal equations}
\label{sec2}
In this paper, we consider the electromagnetic wave propagation
in a heterogeneous, lossless, bi-anisotropic dielectric medium.
We assume that the scattering medium is free of charges and currents and start with 
the source-free Maxwell equations in the frequency  $\om$ domain
\beq
-i\omega\bK\lt[\begin{matrix}
 \bE \\
\bH
\end{matrix}\rt]+\lt[
\begin{matrix}
0 &-\nabla\times\\
\nabla \times &0
\end{matrix}
\rt]\lt[\begin{matrix}
\bE\\
\bH
\end{matrix}
\rt]=0\label{max}
\eeq
where $\bK$ is, by the assumption of losslessness, a Hermitian matrix \cite{LST}
\beq
\label{K}
\bK=\lt[\begin{matrix}
\bK^\epsilon&\bK^\chi\\
\bK^{\chi\dagger}&\bK^\mu
\end{matrix}
\rt]
\eeq
with 
  the permittivity   and 
permeability tensors $\bK^\epsilon, \bK^\mu$, and  the
magneto-electric tensor $\bK^\chi$ \cite{Ode}. 
The Hermitian matrix  $\bK$ is assumed to be always invertible.

In an isotropic dielectric, $\bK^\epsilon=\epsilon\bI, \bK^\mu=\mu\bI, \bK^\chi=0$. In a biisotropic dielectric,
$\bK^\chi$ as well as $\bK^{\epsilon}, \bK^{\mu}$ are nonzero scalars. A reciprocal chiral medium is biisotropic with purely
imaginary $\bK^{\chi}=i\chi $. The appearance of nonzero $\bK^\chi$ arises from
the so called magnetoelectric effect \cite{LLP}.
Crystals are often naturally anisotropic, and in some media (such as liquid crystals) it is possible to induce anisotropy by applying e.g. an external electric field. In crystal optics,  $\bK^\epsilon,\bK^\mu$ are  real, symmetric matrices 
and $\bK^\chi=0$ \cite{BW}. 
In response to a magnetic field, some materials can have a dielectric tensor that is complex-Hermitian; this is called the gyrotropic  effect. In general, a magnetoelectric, bi-anisotropic medium has
a constitutive tensor (\ref{K}) with complex Hermitian $\bK^{\epsilon}, \bK^{\mu}$ and
a complex matrix $\bK^{\chi}$ satisfying
the Post constraint
 \cite{WL}. It has been shown that a moving medium, even isotropic, must be treated as bi-anisotropic
\cite{CK,LLP}.

In general, $\bK$ is a function of the frequency $\om$ 
(for dispersive media) 
but it turns out that if the frequency-dependence of $\bK$ is
sufficiently smooth  the 2f-RT equations derived in
the present framework have the same form as for
nondispersive media;  the frequency-dependence would
enter the coefficients of the equations in the obvious way \cite{2f-rt-josa}. 
For the simplicity of presentation we shall assume 
that the medium is nondispersive.

Writing the total field $\bU=(\bE, \bH)$ we introduce the two-frequency
matrix-valued Wigner distribution
\beq
\label{1.22}
&&\bW(\bx,\bp;\omega_1,\omega_2)\\
&=&\frac{1}{(2\pi)^3}
\int e^{-i\bp^\dagger\by} \bU_1\big(\frac{\bx}{\omega_1}+\frac{\by}{2\omega_1}\big)
 \bU_2^\dagger\big(\frac{\bx}{\omega_2}-\frac{\by}{2\omega_2}\big)d\by\nn
\eeq
where $\bU_1$ and $\bU_2$ are the total fields
at frequencies $\omega_1$ and $ \omega_2$ respectively. 
From the definition we see that the variables $\bx$ and $\bp^{-1}$ have the dimension of $\hbox{length/time}$.  
Although the scaling factors in the arguments of $\bU_{1}$ and
$\bU_{2}$ are not required for the development of 
the 2f-RT theory for the
{\em first-order} (Maxwell's) equations, they are
particularly useful in the case of the {\em second-order} 
(Helmholtz and paraxial wave) equations. For the
 consistency and continuity of presentation (see Section ~\ref{sec7}) we work
 with the definition (\ref{1.22}) in the present 
 paper. For an alternative development of
 the 2f-RT theory for Maxwell's equations in
 terms of the 2f-WD {\em without} the
 scaling factors, we refer  the reader to
 Ref. \cite{2frt-polar}

First note  the symmetry of the Wigner distribution matrix
\beq
\label{4}
\bW^\dagger(\bx,\bp;\omega_1,\omega_2)=\bW(\bx,\bp; \omega_2, \omega_1).
\eeq
 In other words, the right hand side of
(\ref{1.22}) is invariant under the simultaneous  transformations of Hermitian conjugation $\dagger$ and frequency exchange $\omega_1\leftrightarrow \omega_2$. 

In what follows we shall omit writing the arguments of
any fields if there is no risk of confusion. 

We put eq.  (\ref{max})
in the form of general symmetric hyperbolic system \cite{RPK}
\beq
\label{hyper}
-i\om \bK \bU+\mbR_{l}\pxl\bU=0
\eeq
where  the symmetric-matrices  $\mbR_{j}$  are given by
\beqn
\mbR_j=\lt[\begin{matrix}
0&\bT_j\\
-\bT_j&0
\end{matrix}
\rt]
\eeqn
with
\beqn
\bT_1=\lt[\begin{matrix}
0&0&0\\
0&0&-1\\
0&1&0
\end{matrix}
\rt],\quad
\bT_2=\lt[\begin{matrix}
0&0&1\\
0&0&0\\
-1&0&0
\end{matrix}
\rt],\quad\bT_3=\lt[\begin{matrix}
0&-1&0\\
1&0&0\\
0&0&0
\end{matrix}
\rt]. 
\eeqn
The matrices $iT_j, j=1,2,3$ are related to
the photon spin matrices \cite{BB}.

Throughout this paper the dot notation, ``$\cdot$'',   is used exclusively for directional derivative as in $\bp\cdot\nabla=
p_{j}\pxj$.
All vectors are treated as matrices and the scalar product is just the matrix multiplication between row and column vectors. All vectors are taken to be, by default, column vectors,
unless explicitly transposed. 

Applying the operator $\mbR_j\partial/\partial x_j $ to $\bW$ and using (\ref{hyper})  we obtain
\beq
\label{wig1}
&&\mbR_j\frac{\partial}{\partial x_j}\bW\\
&=&-2i  p_j\mbR_j\bW
+2i\int e^{i\bq^\dagger\bx/\omega_1}\widehat \bK(\bq) \bW
\big(\bx, \bp-\frac{\bq}{2\omega_1}\big)d\bq.\nn
\eeq
whose derivation is given in Appendix A.
From (\ref{wig1}) and (\ref{4}) we also have 
\beq
\label{wig2}
&&\frac{\partial}{\partial x_j} \bW\mbR_j^\dagger\\
&=& 2i\bW  p_j\mbR_j-2i
\int  \bW\big(\bx,\bp+\frac{\bq}{2\omega_2}\big) \widehat \bK(\bq)e^{i\bq^\dagger\bx/\omega_2}d\bq.\nn
\eeq
Here and below $\hat\bK$ stands for  the Fourier transform (spectral density)  of  $\bK$ as in 
\[
\bK(\bx)=\int e^{i\bx^\dagger\bq}\widehat \bK(\bq)  d\bq. 
\]
For a Hermitian $\bK$ we have
$
\hat\bK(\bp)=\hat\bK^\dagger(-\bp),\quad\forall \bp.
$

\section{Weak-coupling limit}
\label{sec3}
As in Part I \cite{2f-rt-josa} we consider the weak coupling
regime with the tensor
\beq
\label{weak}
\bK(\bx)=\bK_0\Big(\bI+\sqrt{\ep} \bV\big(\frac{\bx}{\ep}\big)\Big),\quad\ell\ll 1
\eeq
where the  Hermitian matrix $\bK_0$
 represents the uniform background medium
and  $\sqrt{\ep}\bV$ represents the relative fluctuations
of the permittivity-permeability  tensor. The small parameter $\ep$ describes the ratio of
the scale  of the medium
fluctuation to the propagation distance.  
In an isotropic dielectric,
\beqn
\bK_0=\lt[\begin{matrix}
\epsilon_0 \bI_3 &0\\
0&\mu_0\bI_3
\end{matrix}
\rt],\quad
\bV=\lt[\begin{matrix}
\tilde\epsilon \bI_3 &0\\
0&\tilde\mu\bI_3
\end{matrix}
\rt]
\eeqn
where $\tilde\epsilon$ and $\tilde\mu$ are electric and
magnetic susceptibility, respectively. In general $\bK_0$
is 
 a  Hermitian matrix and its blocks, as in (\ref{K}),  are  denoted by
 $\bK_{0}^{\epsilon}, \bK_{0}^{\mu}, \bK_{0}^\chi, \bK_{0}^{\chi\dagger}$,
 respectively. 
To preserve the Hermicity of $\bK$ and $\bK_0$ the matrix  $\bV$ must satisfy
 \beq
 \label{herm}
 \bV^\dagger\bK_0=\bK_0\bV.
 \eeq
  We shall assume below that $\bK_0$ is either positive or negative definite. Otherwise, the materials would be lossy since the refractive index
  is not real-valued if $\bK_0$ is not sign-definite. 
  A negative-definite $\bK_{0}$ gives rise to 
negative  refractive index which is a hot topic in metamaterial research  \cite{Ves,  SPV, SPW}.
To fix the idea, let us take $\bK_{0}$ to be positive definite.
With minor notational change, our method applies equally well to the negative definite case.


We assume that $\bV=[V_{ij}]$  is
a statistically homogeneous random field with
the spectral density tensors ${\mb \Phi}=[\Phi_{ijmn}], {\mb \Psi}=[\Psi_{ijmn}]$ such that
\beq
\lan V_{ij}(\bx)V^*_{mn}(\by)\ran&=&\int e^{i\bk^\dagger(\bx-\by)}\Phi_{ijmn}(\bk)d\bk\\
\lan V_{ij}(\bx)V_{mn}(\by)\ran&=&\int e^{i\bk^\dagger(\bx-\by)}\Psi_{ijmn}(\bk)d\bk.
\eeq
This
 implies the following relations 
\beq
\lan \hat V_{ij}(\bp)\hat V_{mn}^*(\bq)\ran&=&
\Phi_{ijmn}(\bp)\delta(\bp-\bq)\\
\lan \hat V_{ij}(\bp)\hat V_{mn}(\bq)\ran &=&
\Psi_{ijmn}(\bp)\delta(\bp+\bq). 
\eeq

In the case of real-valued $\bV$, ${\mb \Phi}={\mb \Psi}$. 
The spectral density tensors  have 
the basic symmetry
\beq
\label{sym1}
\Phi_{ijmn}^*(\bp)&=&\Phi_{mnij}(\bp),\\
\Psi_{ijmn}(-\bp)&=&\Psi_{mnij}(\bp),
\label{sym2}
\eeq
Furthermore, eq. (\ref{herm}) implies that
\beq
\label{15}
K_{0,ij}\Psi_{mnjl}(\bp)&=&
K^*_{0,lj}\Phi_{mnji}(\bp)\\
K_{0,ij}\Phi_{mnjl}(\bp)&=&K^*_{0,lj}\Psi_{mnji}(\bp)\label{16}
\eeq

As in Part I, we consider the regime
where the wavelengths are of the same order of magnitude as
the correlation length of the medium fluctuations
by rescaling the frequencies $\omega_j= \tilde \omega_j/\ep, j=1,2$.
This choice of frequency scaling results in strong scattering by the
medium heterogeneities.
For ease of notation, we drop the
tilde in $\tilde \omega_j$ below. 
To capture  the high frequency behavior of the wave field
we redefine the 2f-WD as
\beq
\label{1.2}
&&\bW(\bx,\bp)\\
&=&\frac{1}{(2\pi)^3}\nn
\int e^{-i\bp^\dagger\by} \bU_1\big(\frac{\bx}{\omega_1}+\frac{\ell\by}{2\omega_1}\big)
 \bU_2^\dagger\big(\frac{\bx}{\omega_2}-\frac{\ell\by}{2\omega_2}\big)d\by. 
\eeq
We also assume
that $\omega_1, \omega_2\to \om$ as $\ell\to 0$ such that
\beq
\label{shift}
\frac{\omega_2-\omega_1}{\om\ell}=\beta
\eeq
with a fixed  constant  $\beta$.
The governing equations for  (\ref{1.2}) become
\beq
\label{wig3}
\mbR_j\frac{\partial}{\partial x_j}\bW&=&-\frac{2i}{\ep}  p_j\mbR_j \bW
+\frac{2i}{\ep}\bK_0\bW\\
\nn&&+\frac{2i}{\sqrt{\ep}}\int e^{i\bq^\dagger\tilde\bx/\omega_1}\bK_0\widehat \bV(\bq) \bW
\big(\bp-\frac{\bq}{2\omega_1}\big)d\bq\\
\label{wig4}
\frac{\partial}{\partial x_j} \bW \mbR_j
&=& \frac{2i}{\ep}\bW  p_j\mbR_j-\frac{2i}{\ep}
\bW\bK_0\\
\nn&&-\frac{2i}{\sqrt{\ep}}
\int  \bW\big(\bp-\frac{\bq}{2\omega_2}\big) \widehat \bV^\dagger(\bq)\bK_0e^{-i\bq^\dagger\tilde\bx/\omega_2}d\bq
\eeq
where $\tilde\bx=\bx/\ep$ is the fast spatial variable.
In order to cancel the background effect we multiply eq. (\ref{wig3}) by $K_{0}^{-1}$ from left, (\ref{wig4})  by $K_{0}^{-1}$ from right and add them to obtain the symmetrical form
\beq
\label{wig-eq}
&&\bK_0^{-1}\mbR_j\frac{\partial}{\partial x_j}\bW+\frac{\partial}{\partial x_j} \bW\mbR_j\bK_0^{-1}\\
&&+\frac{2i}{\ep}\lt[\bK_0^{-1}  p_j\mbR_j\bW-\bW  p_j\mbR_j\bK_0^{-1}\rt]\nn\\
&=&
\frac{2i}{\sqrt{\ep}}\int \lt[e^{i\bq^\dagger\tilde\bx/\omega_1}\widehat \bV(\bq) \bW
\big(\bp-\frac{\bq}{2\omega_1}\big)\rt.\nn\\
&&\lt. -  \bW\big(\bp-\frac{\bq}{2\omega_2}\big) \widehat \bV^\dagger(\bq)e^{-i\bq^\dagger\tilde\bx/\omega_2}\rt]d\bq.\nn
\eeq
This is the equation that we shall work with to derive the
2f-RT equations emplying the multiscale expansion (MSE) 
\cite{2f-rt-josa, RPK}. Note that eq. (\ref{wig-eq}) 
is invariant under the simultaneous  transformations of Hermitian conjugation $\dagger$ and frequency exchange $\omega_1\leftrightarrow \omega_2$.

If, instead of adding the two equations, we subtract them
then we obtain the anti-symmetric form
\beq
\nn
&&-\frac{4i}{\ell}\bW+\bK_0^{-1}\mbR_j\frac{\partial}{\partial x_j}\bW-\frac{\partial}{\partial x_j} \bW\mbR_j\bK_0^{-1}\\
&&\nn+\frac{2i}{\ep}\lt[\bK_0^{-1}  p_j\mbR_j\bW+\bW  p_j\mbR_j\bK_0^{-1}\rt]\nn\\
&=&
\frac{2i}{\sqrt{\ep}}\int \lt[e^{i\bq^\dagger\tilde\bx/\omega_1}\widehat \bV(\bq) \bW
\big(\bp-\frac{\bq}{2\omega_1}\big)\rt.\nn\\
&&\lt. +\bW\big(\bp-\frac{\bq}{2\omega_2}\big) \widehat \bV^\dagger(\bq)e^{-i\bq^\dagger\tilde\bx/\omega_2}\rt]d\bq.\label{wig-eq2}
\eeq
Eq. (\ref{wig-eq2}) requires a different treatment and
will not be pursued here. However, the leading order $\ell^{-1}$ terms of eq. (\ref{wig-eq2})
impose a constraint which will be discussed in the Conclusion.

\section{Multiscale expansion}
\label{sec4}
The key point of MSE is to separate the fast variable $\tilde\bx$ from
the slow variable $\bx$ and make the substitution 
\beqn
\mbR_j\pxj\bW&\to &
\mbR_j\frac{\partial}{\partial x_j}\bW+
\ep^{-1}\mbR_j\frac{\partial}{\partial \tilde x_j}\bW\\
\pxj\bW\mbR_{j}
&\to& \frac{\partial}{\partial x_j} \bW\mbR_j
+\ep^{-1}\frac{\partial}{\partial \tilde x_j} \bW\mbR_j. 
\eeqn
The idea is that for sufficiently small $\ep$ the two widely separated scales, represented by $\bx $ and $\tilde\bx$ respectively, become  mathematically (but not physically)  independent.

We posit the expansion
$\bW=\bar \bW+\sqrt{\ep} \bW_1+\ep \bW_2+...$, substitute  it  into eq. (\ref{wig-eq}) and equate 
terms of same order of magnitude.

\subsection{Leading term}
The $\ep^{-1}$-terms yield
\beq
\label{eq1}
&&\bK_0^{-1}\mbR_j\frac{\partial}{\partial \tilde x_j}\bar\bW+\frac{\partial}{\partial \tilde x_j} \bar \bW\mbR_j\bK_0^{-1}\\
&&+{2i}\lt[\bK_0^{-1}  p_j\mbR_j\bar \bW-\bar \bW  p_j\mbR_j\bK_0^{-1}\rt]=0. \nn
\eeq
We hypothesize that the leading order term $\bar\bW=\bar\bW(\bx,\bp)$ be
independent of the fast variable $\tilde\bx$. Thus
the first two terms of (\ref{eq1}) vanish
so the equation  reduces to 
\beq
\label{eq2}
\bK_{0}^{-1}  p_j\mbR_j \bar \bW -\bar \bW  p_j\mbR_j
\bK_{0}^{-1}=0. 
\eeq

Eq. (\ref{eq2})  arises also in the equal-time RT theory \cite{RPK} and
can be  solved as follows. 
For a positive (or negative) definite $\bK_{0}$, 
consider the eigenvalues $\{\Om^\sigma\}$ and eigenvectors
$\{\bb^{\sigma, \alpha}\}$ of the
matrix $\bK_{0}^{-1}  p_j\mbR_j$ where the index $\alpha$ keeps track of the multiplicity and hence depends on $\sigma$. As $\bK_{0}^{-1}  p_j\mbR_j$ is
Hermitian  with respect to the scalar product defined by $
{\mb a}^\dagger\bK_{0}{\mb b}, \forall {\mb a}, {\mb b}\in \IC^6$, the eigenvalues are real and
the eigenvectors form a complete set of $\bK_{0}$-orthogonal basis in $\IC^6$. Alternatively,  we may work with the Hermitian 
matrix $\bK_{0}^{{-1/2}}  p_j\mbR_j\bK_{0}^{{-1/2}}$
in the image space,
with the standard  scalar product, 
under the transformation
$\bK^{{1/2}}_{0}$. 
Let the eigenvectors $\{\bb^{\sigma,\alpha}\}$ 
be normalized such that $\bb^{\sigma,\alpha\dagger}\bK_0\bb^{\tau,\zeta}=\delta_{\sigma,\tau}\delta_{\alpha,\zeta}$. 

Clearly, the eigenvalues $\Om^{\sigma}$ as a function of
the wavevector $\bp$ define the dispersion relations. 
For general  bianisotropic dielectric, it is easy to check that  $\Om^0=0$ is
always an eigenvalue with eigenvectors  
\beq
\label{zero}
\bb^{0,1}(\bp)\sim \lt(\begin{matrix}
\bp\\
0
\end{matrix}\rt), \quad
\bb^{0,2}(\bp)\sim \lt(\begin{matrix}
0\\
\bp
\end{matrix}\rt).
\eeq
Since
$\bK_{0}$ is invertible, it follows that
the null space of $\bK_{0}^{-1} p_{j}\mbR_{j}$ is spanned
by these two non-propagating modes. 
It is easy to check that $\{\bd^{\sigma,\alpha\dagger}(\bp): \bd^{\sigma,\alpha}(\bp)=\bK_{0}\bb^{\sigma,\alpha}(\bp)\}$ 
are the {\em left} eigenvectors of $\bK_{0}^{-1}  p_j\mbR_j$
and  $\{\bd^{\sigma, \alpha}(\bp)\}, \{\bb^{\tau,\zeta}(\bp)\}$  are co-orthogonal with respect
to the standard scalar product:
\beq
\label{lr}
\bd^{\sigma,\alpha\dagger}(\bp)\bb^{\tau,\zeta}(\bp)=\delta_{\sigma,\tau}\delta_{\alpha,\zeta}.
\eeq
This relation  will be useful in deriving the 2f-RT equations (see Appendix B). 

Throughout the  English indices represent
the spatial degrees of freedom 
while the Greek indices represent the modal and polarization
degrees of freedom. It is important to keep this
distinction in mind in the subsequent analysis. 
The Einstein summation convention and the Hermitian
conjugation are used only on the English indices.

It can be checked easily that the general
solution to (\ref{eq2}) is given by \cite{RPK}
\beq
\label{w0}
\bar \bW(\bx,\bp)&=&\sum_{\sigma,\alpha,\zeta}\bar W^\sigma_{\alpha\zeta}(\bx,\bp)\bEaz(\bp,\bp)
\eeq
where $\bar W^\sigma_{\alpha\zeta}$ are generally complex-valued functions  and
\beq
\label{right}
\bEaz(\bp,\bq)=\bba(\bp)
\bbz^{\dagger}(\bq).
\eeq
Likewise we define
\beq
\label{left}
\bDaz(\bp,\bq)=\bda(\bp)
\bdz^\dagger(\bq).
\eeq
The linear span of $\{\bE^{\tau,\alpha\zeta}(\bp,\bp),\forall \tau, \alpha,\zeta,\bp\}$ is 
a Hilbert space, denoted by $\fM_\bp$,  for each $\bp\neq 0$ with the scalar product
$\hbox{Tr}\big[\bH^\dagger\bK\bG\bK\big],\bH,\bG\in \fM_\bp$. 
The matrices $\bar \bW^\sigma=[\bar W^\sigma_{\alpha\zeta}]$, free of
the English indices, 
are called the {\em coherence matrices}.

For $\tilde\bx$-independent  $\bar \bW$, the constraint 
that the electric displacement and the magnetic induction 
are both divergence-free yields, on the macroscopic scale, 
\beqn
(\pm\nabla,\pm\nabla)\cdot \bK_{0}\bar\bW=0
\eeqn
which,   in view of 
the definition (\ref{1.2}),  is equivalent to
\beq
(\pm\bp^{\dagger},\pm\bp^{\dagger})\bK_{0}\bar \bW(\bx,\bp)=0.
\eeq
Hence by (\ref{zero}) $\bd^{0,j\dagger} \bar\bW=0$
and by
 (\ref{lr})  $\bar\bW^0=0$ where $\bW^{0}$ in (\ref{w0})  is
 the coherence matrix associated with
 the non-propagating mode $\Om^{0}=0$.

\subsection{Correctors}
The $\ep^{-1/2}$-terms yields the equation
\beq
\nn &&2\ep\bW_1+\bK_0^{-1}\mbR_j\frac{\partial}{\partial \tilde x_j}\bW_1+\frac{\partial}{\partial \tilde x_j} \bW_1\mbR_j\bK_0^{-1}\\
&&+{2i}\lt[\bK_0^{-1}  p_j\mbR_j\bW_1- \bW_1  p_j\mbR_j\bK_0^{-1}\rt]\nn\\
&&={2i}\int d\bq \lt[e^{i\bq^\dagger\tilde\bx/\omega_1}
\widehat\bV(\bq) \bar \bW (\bp-\frac{\bq}{2\omega_1})\rt.\nn\\
&&\lt.
-\bar\bW(\bp-\frac{\bq}{2\omega_2}) \widehat\bV^\dagger(\bq)
e^{-i\bq^\dagger\tilde\bx/\omega_2}\rt]\label{eq3}
\eeq
where, as in Part I\cite{2f-rt-josa}, we have added
a small regularization term. 
The reader is referred to Part I \cite{2f-rt-josa} for
the discussion of the choice of the regularization parameter.
Physically, the sign of the parameter (positive here) amounts
to choosing the direction of causality.

We  Fourier transform eq. (\ref{eq3}) in $\tilde\bx$
\beq
\nn
&&-i2\ep \widehat \bW_1(\bk, \bp)+\bK_0^{-1} k_j\mbR_j\widehat\bW_1(\bk, \bp)+\widehat\bW_1(\bk, \bp) k_j\mbR_j\bK_0^{-1}\nn\\
&&+\nn
{2}\lt[\bK_0^{-1}  p_j\mbR_j\widehat \bW_1(\bk, \bp)-\widehat\bW_1(\bk,\bp)
  p_j\mbR_j\bK_0^{-1}\rt]\\
&=&{2}\lt[\widehat\bV(\omega_1\bk) \bar\bW(\bp-\frac{\bk}{2})
-\bar\bW(\bp+\frac{\bk}{2})\widehat\bV^\dagger(-\omega_2\bk)\rt]
\label{eq4}
\eeq
and posit
the solution 
\beq
\label{w1}
\widehat \bW_1(\bk,\bp)=\sum_{\sigma,\alpha,\zeta}\Caz(\bk, \bp)
\bEaz\big(\bp+\frac{\bk}{2},\bp-\frac{\bk}{2}\big)
\eeq
where $\Caz$ are generally complex numbers. 
Note that the two arguments of $\bEaz$ in (\ref{w1}) are at different
momenta $\bp+{\bk}/{2},\bp-{\bk}/{2}$. 

We substitute (\ref{w0}) and (\ref{w1}) into eq. (\ref{eq4})
and multiply it with $\bda^{\dagger}(\bp+\frac{\bk}{2})$ from
the left and with $\bdz(\bp-\frac{\bk}{2})$ from the right
and solve the resulting equation algebraically. This
 yields the coefficients 
\beq
\nn
&&\Caz(\bk, \bp)\nn\\
&=&\lt(\Om^\sigma(\bp+\frac{\bk}{2})
-\Om^\sigma(\bp-\frac{\bk}{2})-i\ep\rt)^{-1}\nn\\
&&\sum_{\eta}\lt[\bda^{\dagger}(\bp+\frac{\bk}{2})\widehat\bV(\omega_1\bk)\bar W^\sigma_{\eta\zeta}(\bp-\frac{\bk}{2})\bbe(\bp-\frac{\bk}{2})\nn\rt.\\
&&\lt.-\bar W^\sigma_{\alpha\eta}(\bp+\frac{\bk}{2})\bbe^{\dagger}(\bp+\frac{\bk}{2})
\widehat\bV^\dagger(-\omega_2\bk)\bdz(\bp-\frac{\bk}{2})\rt].\label{cij}
\eeq
When  the leading term $\bar \bW$ is invariant
under the simultaneous  transformations of Hermitian conjugation $\dagger$ and frequency exchange $\omega_1\leftrightarrow \omega_2$,
so is $\bW_{1}$. This invariance is manifest in the relation
 \[
 C^{\sigma*}_{\zeta\alpha}(-\bk,\bp; \omega_1, \omega_2)
 =\Caz(\bk, \bp;\omega_2, \omega_1). 
 \]

Finally the $O(1)$-terms yield the equation after adding
a regularizing term $2\ep\bW_2$
\beq
&&2\ep\bW_2+\bK_0^{-1}\mbR_j\frac{\partial}{\partial \tilde x_j}\bW_2+\frac{\partial}{\partial \tilde x_j} \bW_2\mbR_j\bK_0^{-1}\nn\\
&&+{2i}\lt[\bK_0^{-1}  p_j\mbR_j\bW_2- \bW_2  p_j\mbR_j\bK_0^{-1}\rt]=\bF\label{eq5}
\eeq
with
\beq
\bF&=&{2i}\int d\bq \lt[e^{i\bq^\dagger\tilde\bx/\omega_1}
\widehat\bV(\bq) \bW_1 (\bp-\frac{\bq}{2\omega_1})\rt.\nn\\
&&\lt.
-\bW_1(\bp-\frac{\bq}{2\omega_2}) \widehat\bV^\dagger(\bq)
e^{-i\bq^\dagger\tilde\bx/\omega_2}\rt]\nn\\
&&-\bK_0^{-1}\mbR_j \frac{\partial}{\partial x_j}\bar \bW-\frac{\partial}{\partial x_j} \bar \bW \mbR_j\bK_0^{-1}.\label{F}
\eeq
Note again that $\bF$  is invariant under
the simultaneous  transformations of Hermitian conjugation $\dagger$ and frequency exchange $\omega_1\leftrightarrow \omega_2$. 
We can, but need not,  solve eq. (\ref{eq5}) explicitly as
 eq. (\ref{eq4}). However,  in order for  the second perturbation  $\ep\bW_2$  
to vanish in the limit $\ep\to 0$, $\bF$
must satisfy the solvability condition
\beq
\label{sol}
\lim_{\ell\to 0}\hbox{Tr}\lan \bG^\dagger\bK_0\bF\bK_0\ran=0
\eeq
for all random stationary matrices $\bG$ satisfying eq. (\ref{eq1}). This can be seen  by transforming eq. (\ref{eq5}) 
into $\hbox{Tr} \lan \bG^\dagger \bK_0 (\ref{eq5})\bK_0\ran$ which by eq. (\ref{eq1}) implies
$2\ell\hbox{Tr} \lan \bG^\dagger \bK_0 \bW_2\bK_0\ran
=\hbox{Tr} \lan \bG^\dagger\bK_0\bF\bK_0\ran$
and hence (\ref{sol}). 

\commentout{
 Since the left hand side of eq. (\ref{eq5}) at $\ell=0$ is skew-adjoint in the Hilbert space of $\fM_\bp$-valued random stationary processes 
endowed with the inner product defined by the left hand
side of (\ref{sol}), 
$\bF$ must be orthogonal  to the solutions of eq. (\ref{eq1}).
} 

Fortunately, we do not need to work with the full solvability condition (\ref{sol}). It suffices to demand (\ref{sol}) to be fulfilled
by all {\em deterministic} $\bG$, independent
of $\tilde\bx$, such that
\beq
\bK_0^{-1}  p_j\mbR_j\bG- \bG  p_j\mbR_j\bK_0^{-1}=0.
\eeq
In other words, as in (\ref{w0}),  we consider  only a subspace
of the solution space of eq. (\ref{eq1}) and  replace (\ref{sol})  by 
\beq
\label{fred}
\lim_{\ell\to 0}\rm{Tr}\lt(\bD^{\tau, \xi\nu^{\dagger}}(\bp,\bp)
\lan \bF(\bx,\tilde\bx,\bp)\ran\rt)=0,\quad\forall \tau, \xi,\nu, \bp,\bx,\tilde\bx
\eeq
where $\bD^{\tau,\xi\nu}$ are defined in (\ref{left}). 
As noted above, (\ref{wig-eq}), (\ref{eq3})
and (\ref{F}) are invariant under the simultaneous  transformations of Hermitian conjugation $\dagger$ and frequency exchange $\omega_1\leftrightarrow \omega_2$
 and therefore
eq. (\ref{fred}) must also be invariant under the same transformations.

To summarize, we have constructed the three-term expansion $\bar\bW+\sqrt{\ell}\bW_{1}+\ell \bW_{2}$ which is
an approximate solution of the 2f Wigner-Moyal equation
 in
the sense the left hand side of (\ref{wig-eq}) subtracted by
the right hand side of (\ref{wig-eq}) equals exactly
\beqn
&&\sqrt{\ell} \Big[-2\bW_{1}+\bK_{0}^{-1}\mbR_j\pxj
\bW_{1}+\pxj\bW_{1}\mbR_{j}\bK_{0}^{-1}\Big]\\
&&
-2i\sqrt{\ell}\int \lt[e^{i\bq^\dagger\tilde\bx/\omega_1}\widehat \bV(\bq) \bW_{2}
\big(\bp-\frac{\bq}{2\omega_1}\big)\rt.\nn\\
&&\nn\lt. -  \bW_{2}\big(\bp-\frac{\bq}{2\omega_2}\big) \widehat \bV^\dagger(\bq)e^{-i\bq^\dagger\tilde\bx/\omega_2}\rt]d\bq\\
&&
+\ell\Big[-2\bW_{2}+\bK_{0}^{-1}\mbR_j\pxj
\bW_{2}+\pxj\bW_{2}\mbR_{j}\bK_{0}^{-1}
\Big]
\eeqn
which vanishes in a suitable sense as  $\ell\to 0$\cite{2f-rt-josa}.

With (\ref{w1})-(\ref{cij}) and (\ref{F}), eq. (\ref{fred})
is an implicit form of the 2f-RT equations  that determines
the leading order coherence matrix.  Our next step
is to write (\ref{fred}) explicitly in terms of the
spectral densities of the medium fluctuations.  

\section{2f-RT equations}
\label{sec5}

Calculation with the left hand side of eq. (\ref{fred}) is
tedious but straightforward as it involves
only the second order correlations of $\bV$. This
is carried out in Appendix B. 


To state the full result in a concise form, let
us introduce the following quantities. 
Define the  scattering tensors ${\fS}^{\tau}(\bp,\bq)=[\cK^{\tau}_{\xi\nu\alpha\zeta}(\bp,\bq)]$ as
\beq
\label{kernel}
&&\cK^{\tau}_{\xi\alpha\nu\zeta}(\bp,\bq)\\
\nn &= &d^{\tau,\xi*}_s(\bp) e^{\tau, \alpha}_{i}\big(\bq\big)
\Phi_{sifg}\big({\om}(\bp-\bq)\big)d^{\tau,\nu}_f \big(\bp\big) e^{\tau,\zeta*}_{g}\big(\bq\big)
\eeq 
Using (\ref{sym1})-(\ref{16}) one can derive the alternative
expressions for $\cK$:
\beq
&&\cK^{\tau}_{\xi\alpha \nu\zeta}(\bp,\bq) \nn\\
&=&e^{\tau,\xi*}_g(\bp)d^{\tau,\alpha}_f(\bq)
\Psi^*_{fgsi}\big({\om}(\bq-\bp)\big)d^{\tau,\nu}_s(\bp)e^{\tau, \zeta*}_{i}(\bq)\nn\\
 &=&d^{\tau,\xi*}_s(\bp)  e^{\tau, \alpha}_{i}(\bq) \Psi_{fgsi}\big({\om}(\bq-\bp)\big)e^{\tau,\nu}_g(\bp) d^{\tau,\zeta*}_f(\bq) 
. \label{42'}
\eeq
With  (\ref{sym1}), (\ref{sym2}) and (\ref{42'}) it is also straightforward  to check that 
\beq
\label{adj1}
\cK^{\tau*}_{\nu\zeta\xi\alpha}(\bp,\bq)&=\cK^{\tau}_{\xi\alpha\nu\zeta}(\bp,\bq)&=
\cK^{\tau}_{\zeta\nu\alpha\xi}(\bq,\bp)\label{adj2}
\eeq

For any $\fM_\bp$-valued field $\bG(\bp)$ define
the $(\xi,\nu)$-component of the tensor
$\fS^{\tau}(\bp,\bq):\bG(\bq)$ as
\[
\lt[\fS^{\tau}(\bp,\bq):\bG(\bq)\rt]_{\xi\nu}=\sum_{\alpha,\zeta}\cK^{\tau}_{\xi\alpha\nu\zeta}(\bp,\bq)
G_{\alpha\zeta}(\bq).
\]
Define the tensors $\bSig^\tau=[\Sigma^{\tau}_{\xi\nu}]$ analogous to the total scattering cross section 
  as
\beq
\label{total}
\bSig^{\tau}(\bp)&= &\pi\int
\delta\Big( \Om^\tau(\bp)-\Om^\tau(\bq)\Big)\fS^{\tau}
(\bp,\bq):\bI d\bq\\
&&-i\int\cpv \lt(\Om^\tau(\bp)-\Om^\tau(\bq)\rt)^{-1} \fS^{\tau}
(\bp,\bq):\bI d\bq.\nn
\eeq
The  2f-RT equation then reads as
\beq
\label{2frt}
\nabla_{\bp}\Om^{\tau}\cdot
\nabla_{\bx}\bar \bW^{\tau}
&=&2\pi {\om^3}\int \delta\Big(\Om^\tau(\bp)-\Om^\tau(\bq)\Big)\\
&&\times
e^{-i\beta (\bq-\bp)^\dagger\bx}
\fS^{\tau}(\bp,\bq):\bar \bW^\tau(\bq)d\bq\nn\\
&&-{\om^3}\lt(\bSig^{\tau}(\bp)\bar\bW^\tau(\bp)+\bar\bW^\tau(\bp)\bSig^{\tau\dagger}(\bp)\rt),\quad \forall\tau.\nn
\eeq

Introducing the new quantity
\[
\fW^\tau=e^{-i\beta\bp^\dagger\bx}\bar\bW^\tau(\bp)
\]
we recast eq. (\ref{2frt}) into the following form
\beq
\label{final}
&&\nabla_{\bp}\Om^{\tau}\cdot
\nabla_{\bx}\fW^{\tau}+
i\beta\bp\cdot\nabla_{\bp}\Om^{\tau}
\fW^\tau
\\
&=&2\pi\om^3\int \delta\Big(\Om^\tau(\bp)-\Om^\tau(\bq)\Big)
\fS(\bp,\bq):\fW^\tau(\bq) d\bq\nn\\
&&-\om^3\Big[\bSig^\tau(\bp)\fW^\tau(\bp)+\fW^\tau(\bp)\bSig^{\tau\dagger}(\bp)\Big]. \nn
\eeq
This is  the Rayleigh-type scaling behavior  typical
of a random dielectric. 
The cubic, instead of quartic,  power in $\om$ is due to
the appearance of $\om$ as
the scaling factor in the definition of 2f-WD (\ref{1.2}).
The quartic-in-$\om$ law is recovered
upon replacing $\bx$ by $\bx/\om$ on
the left hand side of (\ref{final}).

\subsection{Decoupling: scalar 2f-RT equation}
\label{sec:scalar}
Although,   in view of (\ref{zero}), the zero eigenvalue $\Om^0=0$ has multiplicity
two in general, the nonzero eigenvalues in media other than
the simplest isotropic medium often have multiplicity one
as we shall see in Section~\ref{sec:med}. This is closely
related to the birefringence effect. Under such
circumstances, the 2f-RT equations take a much
simplified form which we now state. 

 Because $\Omega^{j}, j=1,2,3,4$ are simple (multiplicity one), expression (\ref{w0}) reduces to 
 \beqn
\bar \bW(\bx,\bp)&=&\sum_{\sigma}\bar W^\sigma(\bx,\bp)\bE^{\sigma}(\bp,\bp).
\eeqn
In other words, the coherence matrices become scalars
 and
the different polarization modes decouple.
Consequently (\ref{2frt}) becomes
a scalar equation
\beq
\label{final2-scalar}
\nabla_{\bp}\Om^{\tau}\cdot
\nabla_{\bx}\bar W^{\tau} 
&=&{2\pi\om^3}\int \delta\Big(\Om^\tau(\bp)-\Om^\tau(\bq)\Big)\\
&&\times
e^{-i\beta(\bq-\bp)^\dagger\bx}
\fS^\tau(\bp,\bq)\bar W^\tau(\bq) d\bq\nn\\
&&-2{\om^3}\Sigma^\tau(\bp)\bar W^\tau(\bp),\quad \forall\tau\nn
\eeq
where 
\beq
\label{kernel-scalar}
\fS^\tau(\bp,\bq)&=&d^{\tau*}_s(\bp)  e^{\tau}_{i}\big(\bq\big)\Phi_{sifg}\big({\om}(\bp-\bq)\big)d^{\tau}_f\big(\bp\big) 
e^{\tau*}_g\big(\bq\big)\\
\Sigma^\tau(\bp)
&=&\pi\int
\delta\Big( \Om^\tau(\bp)-\Om^\tau(\bq)\Big) \fS^\tau
(\bp,\bq) d\bq.
\label{tot-scalar}
\eeq
Note that the Cauchy singular integral term in (\ref{tot-scalar})
disappears whenever $\Sigma^\tau $ and $\bar\bW$ commute
as in the scalar case. 
From (\ref{final2-scalar})  we can derive
the scalar equation for the quantity 
$\fW^\tau=e^{-i\beta\bp^\dagger\bx}\bar W^\tau(\bp)$
as before.

\section{Special media}
In this section, we consider the eigenstructure of
the dispersion matrix $\bK_0^{-1}p_j\bR_j$ associated with
the various background media for which the scattering 
tensor can be computed explicitly. 
\label{sec:med}
\subsection{Isotropic medium}
\label{sec:iso}
\label{sec6}
In the simplest case of an isotropic medium, 
there are two nonzero  eigenvalues:
$\Om^+(\bp)=c_{0}|\bp|, \Om^-(\bp)=-c_{0}|\bp|$,
each of multiplicity two. 
Let $\hat \bp=\bp/|\bp|$ and let $\hat\bp^+_\perp, \hat\bp^-_\perp$ be any pair of unit vectors orthogonal to each other and to $\hat\bp$ so that $\{\hat\bp^+_\perp, \hat\bp^-_\perp,\hat\bp\}$ form a right-handed coordinate frame. Let $\{\hat\bq^+_\perp, \hat\bq^-_\perp,\hat\bq\}$ 
be similarly defined. The 
eigenvectors are
\beqn
\bb^{+, +}(\bp)=\lt(\begin{matrix}
\frac{1}{\sqrt{2\epsilon_0}}\hat\bp^+_\perp\\
\frac{1}{\sqrt{2\mu_0}}\hat\bp^-_\perp\end{matrix}\rt),\quad
\bb^{+, -}(\bp)=\lt(\begin{matrix}
\frac{1}{\sqrt{2\epsilon_0}}\hat\bp^-_\perp\\
 -\frac{1}{\sqrt{2\mu_0}}\hat\bp^+_\perp\end{matrix}\rt),
\quad\bb^{-, +}(\bp)=\lt(\begin{matrix}
\frac{1}{\sqrt{2\epsilon_0}}\hat\bp^+_\perp\\
 -\frac{1}{\sqrt{2\mu_0}}\hat\bp^-_\perp\end{matrix}\rt),
 \quad
\bb^{-, -}(\bp)=\lt(\begin{matrix}
\frac{1}{\sqrt{2\epsilon_0}}\hat\bp^-_\perp\\
 \frac{1}{\sqrt{2\mu_0}}\hat\bp^+_\perp\end{matrix}\rt). 
\eeqn

Denote the spectral densities of $\tilde\epsilon$ and $\tilde\mu$ by $\Phi_\epsilon$ and $\Phi_\mu$, respectively and denote the cross spectral densities by $\Phi_{\epsilon\mu}, \Phi_{\mu\epsilon}$. 
We have $\fS^\tau=[\cS^\tau_{\xi\nu\alpha\zeta}]$ with
\beq
\cK^\tau_{\xi\alpha\nu\zeta}(\bp,\bq)&=&
\frac{1}{4}
\Big[\Phi_\epsilon({\om}(\bp-\bq))\hat\bp^{\xi\dagger}_\perp\hat\bq^{\alpha}_\perp\hat\bq^{\zeta\dagger}_\perp\hat\bp^{\nu}_\perp\nn\\
&&
-\Phi_{\epsilon\mu}({\om}(\bp-\bq))\hat\bp^{\xi\dagger}_\perp\hat\bq^{\alpha}_\perp\hat\bq^{-\zeta\dagger}_\perp\hat\bp^{-\nu}_\perp\nn\\
&&
-\Phi_{\mu\epsilon}({\om}(\bp-\bq))\hat\bp^{-\xi\dagger}_\perp\hat\bq^{-\alpha}_\perp\hat\bq^{\zeta\dagger}_\perp\hat\bp^{\nu}_\perp\nn\\
&&+
\Phi_\mu({\om}(\bp-\bq))\hat\bp^{-\xi\dagger}_\perp\hat\bq^{-\alpha}_\perp\hat\bq^{-\zeta\dagger}_\perp\hat\bp^{-\nu}_\perp\Big]\label{51'}
\eeq
for $\tau, \xi,\alpha,\zeta,\nu =\pm$.
Eq. (\ref{2frt}) can now be written as
\beq
\label{iso}
c_0\hat\bp\cdot\nabla_\bx \bar\bW^{\pm}
&=&\pm\frac{\pi\om^3|\bp|^2}{4c_{0}}
\Big[2\int e^{-i\beta(\bq-\bp)^\dagger\bx}\delta\Big(|\bp|-|\bq|\Big)\\
&&\times \fS(\bp,\bq):
\bar\bW^{\pm}(\bq)d\hat\bq\nn\\
&&-
\int \delta\Big(|\bp|-|\bq|\Big)\fS(\bp,\bq):\bI d\hat\bq \bar\bW^{\pm}\nn\\
&&
-\bar\bW^{\pm}\int \delta\Big(|\bp|-|\bq|\Big)\fS(\bp,\bq):\bI d\hat\bq 
\Big].\nn
\eeq
The property (\ref{adj1})  and the expression (\ref{51'}) imply that
$\cK^\tau_{\xi\alpha\nu\zeta}(\bp,\bq)
=\cK^\tau_{\xi\alpha\nu\zeta}(\bq,\bp)$ and hence
the Cauchy principal value term in (\ref{total})
disappears.

Often, in a scattering atmosphere for instance, $\tilde\mu=0$ is  a good approximation
and in such case
the only nonzero term in the scattering kernel is 
\beq
\label{chan}
\cK^\tau_{\xi\alpha\nu\zeta}(\bp,\bq)&=&
\frac{1}{4}\Phi_\epsilon({\om}(\bp-\bq))\hat\bp^{\xi\dagger}_\perp\hat\bq^{\alpha}_\perp\hat\bq^{\zeta\dagger}_\perp\hat\bp^{\nu}_\perp
\eeq
This is the  setting
for which S. Chandrasekhar originally derived 
his famous equation of transfer \cite{Cha} 
and 
eq. (\ref{iso}) is just the two-frequency version of
Chandrasekhar's equation. 
\commentout{
To fix the idea,
we  parametrize $\hat\bp, \hat\bp^1_\perp, \hat \bp^2_\perp$ as 
\beq
\label{pvec}
\hat\bp=\lt[\begin{matrix}
\sin{\theta}\cos{\phi}\\
\sin{\theta}\sin{\phi}\\
\cos{\theta}
\end{matrix}
\rt],&\hat\bp^1_\perp=\lt[\begin{matrix}
-\cos{\theta}\cos{\phi}\\
-\cos{\theta}\sin{\phi}\\
\sin{\theta}
\end{matrix}\rt],&\hat\bp^2_\perp
=\lt[\begin{matrix}
-\sin{\phi}\\
\cos{\phi}\\
0
\end{matrix}
\rt]
\eeq
where $\theta$ and $\phi$ are the polar angles
with respect to an appropriately chosen
coordinate system through the point
under consideration. In this representation, the unit vectors $\hat\bp^1_\perp$
and $\hat\bp^2_\perp$ refer to, respectively,  the directions in the meridian
plane and perpendicular to it (see Fig. 8 of Ref. \cite{Cha}). 
Similarly we parametrize $(\hat\bq^1_\perp, \hat\bq^2_\perp,\hat\bq)^\dagger$ with the polar angles $\theta', \phi'$.
Then the Rayleigh phase matrix in Chandrasekhar's transfer equation
can be recovered from (\ref{chan})-(\ref{pvec})
and (\ref{iso}) is just the two-frequency version of
Chandrasekhar's equation. 
}

In the same setting, the new features in (\ref{iso}) beyond
Chandrasekhar's transfer equation  are  the frequency shift $\beta$
and the general form of the power spectrum $\Phi_\epsilon$. In Chandrasekhar's and other cases,
the medium consists of randomly distributed particles of smaller size than the wavelength \cite{Mis, Kok}.
Such a discrete medium 
correspond to a random field $\bV$ 
that is a sum of  $\delta$-like
functions randomly distributed according to
the Poisson point process whose
spectral density tensor $\mb{\Phi}$ can
be calculated. 

\subsection{Chiral media}
\label{sec:chi}
A chiral medium is a reciprocal, biisotropic medium with
 the constitutive matrix 
\[
\bK_{0}=\lt[\begin{matrix}
\epsilon_{0} \bI&i\chi \bI\\
-i\chi\bI& \mu_{0} \bI
\end{matrix}\rt]
\]
where $\chi\in \IR$ is the magneto-electric coefficient.
To maintain a positive-definite $\bK_{0}  $ we assume  $\chi^{2}<\epsilon_0\mu_0$. We then have
\beq
\label{mat}
\bK_{0}^{-1}  p_j\mbR_j=
\frac{c_{0}}{1-\kappa^{2}}\lt[\begin{matrix}
z\bI&-i\kappa\bI\\
i\kappa\bI&z^{{-1}}\bI
\end{matrix}\rt]
\lt[\begin{matrix}
0&-\bp\times\\
\bp\times &0
\end{matrix}
\rt]
\eeq
where $z=\sqrt{\mu_{0}/\epsilon_{0}}>0$ is the impedance and
$ \kappa=\chi c_{0}$ is the chirality parameter.
The  four non-zero simple eigenvalues are
$\Omega^{1}=c_{0}|\bp|(1+\kappa)^{-1},
 \Omega^{2}=c_{0}|\bp|(1-\kappa)^{-1}, 
 \Omega^{3}=c_{0}|\bp|(\kappa-1)^{-1},
 \Omega^{4}=c_{0}|\bp|(-\kappa-1)^{-1}
 $ 
and their  corresponding 
eigenvectors are
\beqn
\bb^{1}\sim \lt(\begin{matrix}
-i\hat\bp^{1}_{\perp}+\hat\bp^{2}_{\perp}\\
-\frac{\hat\bp^{1}_{\perp}}{z}-i\frac{\hat\bp^{2}_{\perp}}{z}
\end{matrix}\rt), 
\bb^{2}\sim  \lt(\begin{matrix}
i\hat\bp^{1}_{\perp}+\hat\bp^{2}_{\perp}\\
-\frac{\hat\bp^{1}_{\perp}}{z}+i\frac{\hat\bp^{2}_{\perp}}{z}
\end{matrix}\rt), 
\bb^{3}\sim \lt(\begin{matrix}
-i\hat\bp^{1}_{\perp}+\hat\bp^{2}_{\perp}\\
\frac{\hat\bp^{1}_{\perp}}{z}+i\frac{\hat\bp^{2}_{\perp}}{z}
\end{matrix}\rt),
\bb^{4}\sim \lt(\begin{matrix}
i\hat\bp^{1}_{\perp}+\hat\bp^{2}_{\perp}\\
\frac{\hat\bp^{1}_{\perp}}{z}-i\frac{\hat\bp^{2}_{\perp}}{z}
\end{matrix}\rt). 
\eeqn
Note also that $\Om^{4}=-\Om^{1}, \Om^{3}=-\Om^{2}$. 
As $|\kappa|<1$ (since $\chi^2<\epsilon_0\mu_0$), $\bb^{1}, \bb^{2}$ are the forward propagating modes and $\bb^{3},\bb^{4}$ the backward
propagating modes. 

 For the medium fluctuation $\bV$ we may use the
 biisotropy form 
 \beqn
 \lt[
 \begin{matrix}
 a\bI& i\mu_{0} b \bI\\
 -i\epsilon_{0} b \bI& a\bI
 \end{matrix}
 \rt]
 \eeqn
 where $a, b\in \IR$ are stationary random functions of $\bx$
 with power spectral densities $\Phi_{a}, \Phi_{b}$ and
 the cross-spectral density $\Phi_{ab}$. 
 This particular form is derived from the commutativity relation
 (\ref{herm}).

 The splitting into two distinct positive dispersion relations
 is a case of birefrigence where two distinct phase
 velocities, $c_{0}/(1\pm \kappa)$, arise depending on the polarization. 
As discussed in Section \ref{sec:scalar} due to the birefringence the chiral  medium does not depolarize 
the electromagnetic waves. 
 For the sake of space, we leave
 to the reader to work out  the scattering
 tensor from (\ref{kernel-scalar})-(\ref{tot-scalar}).

 \subsection{Birefrigence in anisotropic crystals}
 \label{sec:aniso}
 Generally speaking, an anisotropic medium permits
 two monochromatic plane waves with two different
 linear polarizations and two different 
 velocities to propagate in any given
 direction \cite{BW}. This again gives rise to the birefringence effect.
 
 The only optically isotropic crystal is the
 cubic crystal. 
 In the system of principal dielectric axes, the permitivity-permeability  tensor
 of a crystal, which is always a real, symmetric matrix,  can be diagonalized as $\bK_{0}=\hbox{diag}[\epsilon_{x}, \epsilon_{y}, \epsilon_{z}, 1,1,1]$. One  type of anisotropic crystals  are  the uniaxial crystals for which $\epsilon_{x}=\epsilon_{y}=\epsilon_{\perp}\neq \epsilon_{z}=\epsilon_{\parallel}$ (if the distinguished
 direction, the optic axis,  is taken as the $z$-axis). 
  There exist two distinct
 dispersion relations for the forward modes
 \beqn
\Om^{o}&=\frac{|\bp|}{\sqrt{\epsilon_{\perp}}},\quad
 \Omega^{e}&=\sqrt{\frac{p_{3}^{2}}{\epsilon_{\perp}}+
 \frac{p_{1}^{2}+p_{2}^{2}}{\epsilon_{\parallel}}}.
 \eeqn
 The backward modes correspond to $-\Om^{0}, -\Om^{e}$. 
The corresponding  wavevector surface consists of 
 a sphere  and an ovaloid, a surface
 of revolution. $\Om^o$ 
 corresponds to the {\em ordinary} wave with a velocity independent
 of the wavevector while $\Om^e$ corresponds to  the
 {\em extraordinary} wave with a velocity depending on the angle
 between the wavevector and the optic axis \cite{BW}. 

Let  $\bd^{o}, \bd^{e}$ be the associated left
 eigenvectors. 
Set
 $\bK_{0}^{\epsilon}=\hbox{diag}[\epsilon_{\perp},
 \epsilon_{\perp},\epsilon_{\parallel}]$ and let
 ${\mb a}^{\sigma}$ solve the following symmetric eigenvalue problem:
\beq
\label{eig2}
-\bp\times \big(\bK_{0}^{\epsilon}\big)^{-1}\bp\times
{\mb a}^{\sigma}=\big(\Om^{\sigma}\big)^{2}{\mb a}^{\sigma},\quad \sigma=e, o.
\eeq
Then the left eigenvectors
  $\bd^{\sigma}$ can be written as
 \beq
 \label{eig1}
 \bd^{\sigma}&\sim &\lt(\begin{matrix}
 -\bp\times
 {\mb a}^{\sigma}\\
  \Om^{\sigma} {\mb a}^{\sigma}
  \end{matrix}\rt),\quad\sigma= e, o.
 \eeq
 The same formula applies to the backward modes. 
 Eq. (\ref{eig2}) has the following solutions
  \[
 {\mb a}^{e}=(-p_{2}, p_{1}, 0)^{\dagger},\quad
 {\mb a}^{o}=(p_{1}, p_{2}, -\frac{p_{1}^{2}+p_{2}^{2}}{p_{3}})^{\dagger}
 \]
 from which we see that the wave is linearly polarized. 
 
The other type of anisotropic crystals is  the  biaxial crystals
for which  there are also 
 two distinct, but more complicated,  dispersion relations,
  both associated with the 
 extraordinary waves \cite{BW}. In contrast, the two
 distinct dispersion relations of a chiral medium  give rise to 
two ordinary waves as the two wavevector surface consists
of two concentric spheres centered at $\bp=0$. 
 
 It should be emphasized that a plane wave propagating
 in an anisotropic crystal is  linearly polarized in certain planes whereas a plane wave propagating in the isotropic medium
 is in general elliptically polarized, and is linearly polarized
 only in particular cases. 
 In the anisotropic as well as  the chiral media,  the different polarizations decouple
 in the radiative transfer equations and 
the depolarization effect is absent. 

\subsection{Gyrotropic media}
\label{sec:gyro}
In the presence of a static external magnetic field $\bH_{\rm ext}$ the permittivity tensor $\bK^{\epsilon}_{0}$  is no longer symmetrical; it is
generally a complex Hermitian matrix. Here we consider
the simplest such constitutive relation
\beq
\label{gyrok}
\bD=\epsilon_{0}\bE-i\bg\times\bE,\quad\bB=\bH
\eeq
where  ${\mb g}=f\bH_{\rm ext}, f\in \IR, $ is the gyration vector. Equivalently, we can write 
\[
\bE=\frac{1}{\epsilon_{0}^{2}-|\bg|^{2}}
\lt(\epsilon_{0}\bD+i\bg \times \bD-\frac{1}{\epsilon_{0}}\bg\bg^{\dagger}\bD\rt). 
\]
In this case there are two distinct  forward dispersion
relations \cite{LLP}
\[
{\Om^{1}}=c_{0}\big|\bp+\frac{\Om^{1}}{2}\bg\big|,
\quad {\Om^{2}}=c_{0}\big|\bp-\frac{\Om^{2}}{2}\bg\big|
\]
 where $c_{0}=1/\sqrt{\epsilon_{0}}$.  
Clearly the wave-vector surface consists of two spheres of the same radius  but different centers. This should be contrasted with
the case of chiral media for which the wave-vector surface consists of two concentric spheres of different radii. 

The associated (left) eigenvectors $\bd^{\sigma}, \sigma=1,2$ can be written as in (\ref{eig1})
with ${\mb a}^{\sigma}$ solving 
(\ref{eig2}) and  $\bK_{0}^{\epsilon}$ corresponding to 
(\ref{gyrok}). 
Let $\bg=g_{1}\hat\bp^{1}_{\perp}+g_{2}\hat\bp^{2}_{\perp}
+g_{3}\hat\bp$. We can write the three-dimensional vector ${\mb a}^{\sigma}$ as ${\mb a}^{\sigma}=\hat\bp^{1}_{\perp}+\gamma_{\sigma}\hat\bp^{2}_{\perp}$ with 
\beqn
\gamma_{\sigma}=\frac{g_{2}^{2}-g_{1}^{2}
-(-1)^{\sigma}\sqrt{(g_{1}^{2}+g_{2}^{2})^{2}+4\epsilon_{0}^{2}g_{3}^{2}}}{2(g_{1}g_{2}-i\epsilon_{0}g_{3})},\quad \sigma=1,2.
\eeqn
We see that
the wave is in general elliptically  polarized or linearly polarized  when $\bg$ is orthogonal to the wavevector $\bp$ and circularly polarized when
$\bg$ is parallel to $\bp$.  
Again,  the simplicity of the eigenvalues implies
that depolarization is absent in
 the gyrotropic media.

\section{Geometrical 2f-RT}
\label{sec7}
We have seen in Section \ref{sec:scalar} how a scalar
2f-RT equation naturally arises  in a birefringent medium. 
In this section, we show that a scalar 2f-RT equation can also
arise in a depolarizing medium such as the isotropic medium
discussed in Section \ref{sec:iso}.
Depolarization can mix different polarization modes
and result in scalar-like
coherence matrices $\bar \bW^\tau \approx \bar W^{\tau} \bI, \tau=\pm$
(see Section \ref{sec:iso} for notation). The other purpose
of this Section is to show smooth transition from
(\ref{2frt}) to the Fokker-Planck equation, previously derived  for the scalar waves \cite{2f-rt-josa, 2f-grt}, 
in the geometrical optics through rapid depolarization. 

Let us start with the general setting and replace  $ 
{\mb \Phi}(\cdot)$ in (\ref{final}) by
$\ga^{{-4}}{\mb \Phi}(\cdot/\ga)$
where  the small parameter  $\ga$ is roughly the ratio
of the wavelength to the correlation length of the medium
fluctuations. In other words, we consider the geometrical optics regime. The quartic power in $\ga$ is  indicative
of the Rayleigh-type scattering. 
Consider  the change of variable
$\bq=\bp+\ga\bk$ in the scattering term of
(\ref{final}). With this  and  the ansatz $\fW^{\tau}=e^{-i\beta\bp\dagger\bx}\bar W^{\tau}$ the scattering term  
becomes approximately 
\beq
\nn
&&2\pi\om^{3}\ga^{{-1}}\int d\bk \delta\Big(\Om^{\tau}(\bp+\ga\bk)-\Om^{\tau}(\bp)\Big)d^{\tau,\xi*}_s(\bp) d^{\tau,\nu}_f\big(\bp\big) \\
&&\nn\times
\Phi_{sifg}\big({\om}\bk\big)\sum_{{\alpha}}e^{\tau, \alpha}_{i}\big(\bp+\ga\bk\big)e^{\tau,\alpha*}_g\big(\bp+\ga\bk\big)\\
&&\times\Big[\fW^\tau(\bp)+
\ga\bk\cdot\nabla_{\bp} \fW^{\tau}(\bp) 
+\frac{\ga^{2}}{2}
k_{l}k_{j}\partial_{p_{l}}\partial_{p_{j}}
\fW^{\tau}(\bp)\Big]
\label{kernel2}.
\eeq
The first term in (\ref{kernel2}) cancels exactly with 
$\bSig^\tau(\bp)\fW^\tau(\bp)+\fW^\tau(\bp)\bSig^{\tau\dagger}(\bp)$ on the right hand side
of (\ref{final}). The second term in (\ref{kernel2})  yields the first order differential operator 
\beq
\label{50.1}
&&\Big[\pi\om^{3}d^{\tau,\xi*}_s d^{\tau,\nu}_f\sum_{{\alpha}}e^{\tau, \alpha}_{i}e^{\tau,\alpha*}_g[\partial_{p_{l}}\partial_{p_{j}}\Om^{\tau}]\\
&&\times\nn\int 
k_{l}k_{j}\delta'\Big(\bk\cdot\nabla_{\bp}\Om^{\tau}\Big)
\Phi_{sifg}\big({\om}\bk\big) \bk d\bk 
\\
&&+2\pi\om^{3}d^{\tau,\xi*}_sd^{\tau,\nu}_f \sum_{{\alpha}}\partial_{p_{l}}\big[e^{\tau, \alpha}_{i}e^{\tau,\alpha*}_g\big]\nn\\
&&\times\int k_{l}
\delta\Big(\bk\cdot\nabla_{\bp}\Om^{\tau}\Big)
\Phi_{sifg}\big({\om}\bk\big)
\bk d\bk \Big]\cdot\nabla_{\bp}\fW^{\tau}\nn
\eeq
where $\delta'$ is the derivative of the Dirac-delta function.
And the third term in (\ref{kernel2}) yields
the second order differential operator
\beq
\nn
&&\pi\om^{3}d^{\tau,\xi*}_s d^{\tau,\nu}_f \sum_{{\alpha}}e^{\tau, \alpha}_{i}e^{\tau,\alpha*}_g\int 
\delta\Big(\bk\cdot\nabla_{\bp}\Om^{\tau}\Big)
 \\
&&\times\Phi_{sifg}\big({\om}\bk\big)k_{m}k_{n}d\bk
\partial_{p_{m}}\partial_{p_{n}}
\fW^{\tau}.\label{50.2}
\eeq

In order to match the left hand side of (\ref{final}) which
is a scalar in case of complete depolarization,  (\ref{50.2}) and each term in
(\ref{50.1}) must be proportional to
$\delta_{\xi,\nu}$ as well. 
This happens, for instance, for the  isotropic medium with (\ref{chan}).
In this case,
\beqn
(\ref{50.1})&=&\pi\delta_{{\xi,\nu}}
\frac{\om^{3}}{4}\int\bk\cdot\nabla_{\bp}(\bk\cdot\hat\bp)
\delta'\Big(\bk\cdot\hat \bp\Big)
\Phi_{\epsilon}\big({\om}\bk\big) \bk d\bk \cdot\nabla_{\bp}\fW^{\tau},\\
(\ref{50.2})&=&\pi
\delta_{{\xi,\nu}}\frac{\om^{3}}{4c_{0}}\int 
\delta\Big(\bk\cdot\hat\bp\Big)
\Phi_{\epsilon}\big({\om}\bk\big)k_{m}k_{n}d\bk 
\partial_{p_{m}}\partial_{p_{n}}
\fW^{\tau}\nn
\eeqn
and hence the 2f-RT equation (\ref{final}) becomes
\beqn
\pm c_{0}\hat\bp\cdot
\nabla_{\bx}\fW^{\pm}\pm
i\beta  c_{0}|\bp|
\fW^\pm
&=&
\nabla_{\bp}\cdot \bD\nabla_{\bp}\fW^{\pm}
\eeqn
with  the diffusion coefficient
\beq
\label{66}
\bD=\frac{\pi\om^{3}}{4c_{0}}
\int 
\delta\Big(\bk\cdot\hat\bp\Big)
\Phi_{\epsilon}\big({\om}\bk\big)\bk \bk^\dagger d\bk 
\eeq
which is the same  Fokker-Planck-type equation 
derived by a rigorous, probabilistic method  from the geometrical optics of the scalar wave
previously \cite{2f-grt}.

Applying  the same procedure to the scalar 2f-RT equation 
for the quantity
$\fW^{\tau}$ of the birefringence case  discussed in Section \ref{sec:scalar}, we obtain
\beqn
\nabla_{\bp}\Om^{\tau}\cdot\nabla_{\bx}\fW^{\tau}
+i\beta\bp\cdot\nabla_{\bp}\Om^{\tau} \fW^{\tau}
=\nabla_{\bp}\cdot \bD\nabla_{\bp} \fW^{\tau}
\eeqn
where  the diffusion coefficient $\bD$ is given by
\beqn
\bD(\bp)&=& \pi {\om^{3}}d^{\tau*}_s(\bp) e^{\tau}_{i}\big(\bp\big)d^{\tau}_f\big(\bp\big) e^{\tau*}_g\big(\bp\big) \\
&&\times \int 
\delta\Big(\bk\cdot\hat\bp\Big)
\Phi_{sifg}\big({\om}\bk\big)
\bk \bk^\dagger d\bk . 
\eeqn 

\section{Conclusion}
\label{sec:conc}
Starting with the symmetrical Wigner-Moyal
equation  (\ref{wig-eq}), we have systematically derived the 2f-RT equations
(\ref{2frt}), (\ref{final}) and (\ref{final2-scalar}) in the radiative transfer regime characterized by  the weak coupling
scaling (\ref{weak}).  The main assumptions on the medium are
that the background is {\em uniform} and has a either positive or negative definite constitutive matrix and that  the fluctuations are zero-mean statistically homogeneous processes.

We not turn to the antisymmetrical Wigner-Moyal equation 
(\ref{wig-eq2}) and discuss the consequence of its
leading order terms which are
\[
2\bar\bW=\bK_0^{-1}  p_j\mbR_j\bar\bW+\bar\bW  p_j\mbR_j\bK_0^{-1}.
\]
In view of (\ref{w0}) this is equivalent to
$1=\Om^{\sigma}(\bp)$. Note again the variable $\bp$
has the dimension of inverse velocity. 
Therefore 
the two-spacetime correlations
of the time dependent  polarized wave field $\bu$ 
are given approximately by 
\beq
\label{two}
&&\lan \bu(t_1,\bx_1)  \bu^\dagger(t_2,\bx_2)\ran\\
&\sim &\sum_{\tau,\alpha\zeta}\int\int e^{i \omega \beta t}e^{-i\omega\tau /\ell}\int_{\Om^{\sigma}(\bp)=1}   e^{i\bp^\dagger(\by+\beta\bx)}  \fW^{\tau}_{\alpha\zeta}(\bx,\bp)\bE^{\tau,\alpha\zeta}(\bp,\bp)
d\Om(\bp) d\omega d\beta,\nn
\eeq
with $\bx=\omega(\bx_{1}+\bx_{2})/2, \by=
\omega(\bx_{1}-\bx_{2})/\ep$
where $\fW^\tau=[\fW^\tau_{\alpha\zeta}]$ is the solution to eq. (\ref{final})
and $d\Om(\bp)$ is the area element of the surface $\Om^{\sigma}(\bp)=1$. 
 
Parallel to the case of scalar waves one can also work
out the 
implications of the polarization on the problems of imaging and
time-reversal communications, as discussed in the Introduction and references therein, from
the 2f-RT theory developed here.

\section{Acknowledgements}

Research is supported by ONR Grant N00014-02-1-0090, Darpa Grant 
 N00014-02-1-0603.

\begin{appendix}
\hspace{-2cm}
\section{Derivation of Wigner-Moyal equation}
Applying the operator $\mbR_j\partial/\partial x_j$ to $\bW$  we have
\beqn
&&\mbR_j\frac{\partial}{\partial x_j}\bW(\bx,\bp)\\
&=&\frac{1}{(2\pi)^3}
\int e^{-i\bp^\dagger\by}
\mbR_j\frac{\partial}{\partial x_j} \bU_1\big(\frac{\bx}{\omega_1}+\frac{\by}{2\omega_1}\big) \bU^\dagger_2
\big(\frac{\bx}{\omega_2}-\frac{\by}{2\omega_2}\big)d\by\\
&&
+\frac{1}{(2\pi)^3}
\int e^{-i\bp^\dagger\by}
\mbR_j\bU_1\big(\frac{\bx}{\omega_1}+\frac{\by}{2\omega_1}\big)  \frac{\partial}{\partial x_j}\bU^\dagger_2
\big(\frac{\bx}{\omega_2}-\frac{\by}{2\omega_2}\big)
d\by
\eeqn
\beqn
&=&\frac{1}{(2\pi)^3}
\int e^{-i\bp^\dagger\by}
\omega_1^{-1}\mbR_{j}\frac{\partial}{\partial x_{j}}\bU_1\big(\frac{\bx}{\omega_1}+\frac{\by}{2\omega_1}\big) \bU^\dagger_2
\big(\frac{\bx}{\omega_2}-\frac{\by}{2\omega_2}\big)d\by\\
&&
-\frac{2}{(2\pi)^3}
\int e^{-i\bp^\dagger\by}
\mbR_j\bU_1\big(\frac{\bx}{\omega_1}+\frac{\by}{2\omega_1}\big) \frac{\partial}{\partial y_j}\bU^\dagger_2
\big(\frac{\bx}{\omega_2}-\frac{\by}{2\omega_2}\big)d\by.
\eeqn
Integrating by parts with the second integral and using (\ref{hyper}) we obtain
\beqn
&&\mbR_j\frac{\partial}{\partial x_j}\bW(\bx,\bp)\\
&=&
\frac{2i}{(2\pi)^3}
\int e^{-i\bp^\dagger\by}
\bK\big(\frac{\bx}{\omega_1}+\frac{\by}{2\omega_2}\big)\bU_1\big(\frac{\bx}{\omega_1}+\frac{\by}{2\omega_1}\big) \bU^\dagger_2
\big(\frac{\bx}{\omega_2}-\frac{\by}{2\omega_2}\big)d\by\\
&&
-\frac{2i}{(2\pi)^3}  p_j\mbR_j
\int e^{-i\bp^\dagger\by}
\bU_1\big(\frac{\bx}{\omega_1}+\frac{\by}{2\omega_1}\big) \bU^\dagger_2
\big(\frac{\bx}{\omega_2}-\frac{\by}{2\omega_2}\big)d\by.
\eeqn
Inserting the spectral representation of $\bK$ 
intro the equation and using the definition (\ref{1.2}) we
then obtain (\ref{wig1}). 
\section{Calculation of eq. (\ref{fred})}
\subsection{Propagation terms}
We first show that
\[
\hbox{Tr}\Big[\bD^{\tau,\xi\nu\dagger}\big(\bK_0^{-1}\mbR_j\pxj\bar \bW+\pxj\bar \bW \mbR_{j}\bK_0^{-1}\big)\Big]
=2\nabla_{\bp}\Om^{\tau}\cdot\nabla_{\bx} \bar W^{\tau}_{\xi\nu}.
\]
Consider the following calculation
\beqn
&&\bK_0^{-1}\mbR_j\pxj\bar \bW\\
&=&
\nabla_{\bp}\big[\bK_{0}^{-1}  p_j\mbR_j\big]\cdot
\big[\nabla_{\bx}\bar W^{\sigma}_{\alpha\zeta}\big]\bb^{\sigma,\alpha}
\bb^{\sigma,\zeta\dagger}\\
&=&\nabla_{\bp}\big[\bK_{0}^{-1}  p_j\mbR_j\bb^{\sigma,\alpha}\big]\cdot
\big[\nabla_{\bx}\bar W^{\sigma}_{\alpha\zeta}\big]
\bb^{\sigma,\zeta\dagger}\\
&&-
\bK_{0}^{-1}  p_j\mbR_j\big[\nabla_{\bp}\bb^{\sigma,\alpha}\big]\cdot
\big[\nabla_{\bx}\bar W^{\sigma}_{\alpha\zeta}\big]
\bb^{\sigma,\zeta\dagger}\\
&=& \nabla_{\bp}\Om^{\sigma}\cdot\nabla_{\bx}
\bar W^{\sigma}_{\alpha\zeta}\bb^{\sigma,\alpha}
\bb^{\sigma,\zeta\dagger}\\
&&
+\Big(\Om^{\sigma}-\bK^{-1}_{0}  p_j\mbR_j\Big)
\big[\nabla_{\bp}\bb^{\sigma,\alpha}\big]\cdot
\big[\nabla_{\bx}\bar W^{\sigma}_{\alpha\zeta}\big]
\bb^{\sigma,\zeta\dagger}.
\eeqn
 Upon the operation $\hbox{Tr}
\big[\bD^{\tau,\xi\nu}\big(\cdot\big)\big]$ the second term vanishes
while the first term reduces to
$\nabla_{\bp}\Om^{\tau}\cdot\nabla_{\bx}
\bar W^{\tau}_{\xi\nu}$ by (\ref{lr}) and the fact that
$\bd^{\tau,\xi}$ is a left eigenvector of the matrix $\bK_{0}^{-1}  p_j\mbR_j$ with the eigenvalue $\Om^{\tau}$. 

The other term $\hbox{Tr}\big[\bD^{\tau,\xi\nu\dagger}\pxj\bar \bW \mbR_{j}\bK_0^{-1}\big]$ gives the identical result.

\subsection{Scattering kernel}
The $(s,j)$-element of the matrix
\beqn
&&\lan \bF\ran+\bK_0^{-1}\mbR_j \frac{\partial}{\partial x_j}\bar \bW+\frac{\partial}{\partial x_j} \bar \bW \mbR_j\bK_0^{-1}\\
&=&{2i}\int d\bq \lan e^{i\bq^\dagger\tilde\bx/\omega_1}
\widehat\bV(\bq) \bW_1 (\bp-\frac{\bq}{2\omega_1})\rt.\nn\\
&&\lt.
-\bW_1(\bp-\frac{\bq}{2\omega_2}) \widehat\bV^\dagger(\bq)
e^{-i\bq^\dagger\tilde\bx/\omega_2}\ran\nn
\eeqn
has  the expression 
\beqn
&&\sum_{\sigma,\alpha,\zeta,\eta} {2i\omega_1^3}\int d\bk \lt(\Om^\sigma(\bp+\bk)-\Om^\sigma(\bp)-i\ell\rt)^{-1}\nn\\
&&\times
d^{\sigma,\alpha*}_f(\bp+\bk)\Psi_{fgsi}(\omega_1\bk)
\bar W^\sigma_{\eta\zeta}(\bp) e^{\sigma,\eta}_g(\bp)
E^{\sigma, \alpha\zeta}_{ij}(\bp+\bk,\bp)\nn\\
&&-{2i\omega_1^3}\int d\bk \lt(\Om^\sigma\big(\bp+\frac{1}{2}(\frac{\omega_2}{\omega_1}+1)\bk\big)-\Om^\sigma\big(\bp+\frac{1}{2}(\frac{\omega_2}{\omega_1}-1)\bk\big)-i\ell\rt)^{-1}\nn\\
&&\times e^{i(1-\frac{\omega_2}{\omega_1})\bk^\dagger\tilde\bx}
\bar W^\sigma_{\alpha\eta} 
\big(\bp+\frac{1}{2}(1+\frac{\omega_2}{\omega_1})\bk\big)\Phi^*_{fgsi}(-\omega_2\bk) \\
&& \times e^{\sigma,\eta*}_g\big(\bp+\frac{1}{2}(1+\frac{\omega_2}{\omega_1})\bk\big)
d^{\sigma,\zeta}_f\big(\bp+\frac{1}{2}(\frac{\omega_2}{\omega_1}-1)\bk\big)\nn\\
&&\times
E^{\sigma, \alpha\zeta}_{ij}\big(\bp+\frac{1}{2}(1+\frac{\omega_2}{\omega_1})\bk,\bp
+\frac{1}{2}(\frac{\omega_2}{\omega_1}-1)\bk\big)\nn\\
&&-{2i\omega_2^3}\int d\bk 
\lt(\Om^\sigma\big(\bp+\frac{1}{2}(1-\frac{\omega_1}{\omega_2})\bk\big)-\Om^\sigma\big(\bp-\frac{1}{2}(\frac{\omega_1}{\omega_2}+1)\bk\big)-i\ell\rt)^{-1}\nn\\
&&\times e^{i(1-\frac{\omega_1}{\omega_2})\bk^\dagger\tilde\bx} 
\bar W^\sigma_{\eta,\zeta} \big(\bp-\frac{1}{2}(\frac{\omega_1}{\omega_2}+1)\bk\big)\Phi_{fgjn}(\omega_1\bk)\\
&&\times d^{\sigma,\alpha*}_f \big(\bp+\frac{1}{2}(1-\frac{\omega_1}{\omega_2})\bk\big) e^{\sigma,\eta}_g
\big(\bp-\frac{1}{2}(\frac{\omega_1}{\omega_2}+1)\bk\big)\nn\\
&&\times
E^{\sigma,\alpha\zeta}_{sn}
\big(\bp+\frac{1}{2}(1-\frac{\omega_1}{\omega_2})\bk, \bp-\frac{1}{2}(\frac{\omega_1}{\omega_2}+1)\bk\big)\nn\\
&&+{2i\omega_2^3}\int d\bk \lt(\Om^\sigma(\bp)-\Om^\sigma(\bp-\bk)-i\ell\rt)^{-1}\nn\bar W^\sigma_{\alpha\eta}(\bp)\Psi^*_{fgjn}(-\omega_2\bk)\\
&&\times 
e^{\sigma,\eta*}_g(\bp)
 d^{\sigma,\zeta}_f(\bp-\bk)
E^{\sigma, \alpha\zeta}_{sn}(\bp,\bp-\bk).\nn
\eeqn
Using the identity 
\beqn
\label{cpv}
\lim_{\ell\to 0}\frac{1}{x -i\ell}=i\pi \delta(x)+\frac{1}{x},
\eeqn
the symmetry properties (\ref{sym1})-(\ref{sym2})  and
(\ref{zero}) we obtain 
in the limit $\ell\to 0$ eq. (\ref{2frt}) from
(\ref{fred}). 
\end{appendix}

\end{document}